\documentclass[final,3p,times]{elsarticle}

\usepackage{lineno,float,amsmath,hyperref}
\usepackage[font=normalsize]{caption}
\usepackage{listings}
\usepackage{makecell}
\usepackage{picins}

\usepackage{setspace}
\modulolinenumbers[5]

\journal{Computers \& Electrical Engineering}

\biboptions{sort&compress}

\begin{document}

\begin{frontmatter}



\title{A practical approach to detection of distributed denial-of-service attacks using a hybrid detection method}


\author[pb]{P.D. Bojovic\corref{cor1}}
\cortext[cor1]{P.D. Bojovic}
\ead{petar.bojovic@paxy.in.rs}
\author[ib]{I. Basicevic}
\author[so]{S. Ocovaj}
\author[ib]{M. Popovic}
\address[pb]{School of Computing University Union Belgrade, 6/6 Knez Mihailova, Belgrade, Serbia }
\address[ib]{Faculty of Technical Sciences University of Novi Sad, 6 Trg Dositeja Obradovića, 
Novi Sad, Serbia}
\address[so]{RT-RK Institute for Computer Based Systems, 23a Narodnog Fronta, Novi Sad, Serbia}

\begin{abstract}
This paper presents a hybrid method for the detection of distributed denial-of-service (DDoS) attacks that combines feature-based and volume-based detection. Our approach is based on an exponential moving average algorithm for decision-making, applied to both entropy and packet number time series. The approach has been tested by performing a controlled DDoS experiment in a real academic network. The network setup and test scenarios including both high-rate and low-rate attacks are described in the paper. The performance of the proposed method is compared to the performance of two methods that are already known in the literature. One is based on the counting of SYN packets and is used for detection of SYN flood attacks, while the other is based on a CUSUM algorithm applied to the entropy time series. The results show the advantage of our approach compared to methods that are based on either entropy or number of packets only.
\end{abstract}


\end{frontmatter}


\section{Introduction}
\label{sec:Intro}





Modern technological society is greatly dependent on Internet technology and online services. Internet services have become a non-exclusive part of everyday routine. Many of us check our e-mail as the first thing we do in the morning. This kind of service dependence has made room for a new kind of manipulation and has introduced attacks on network services. Denial of Service (DoS) attacks are among these attacks. Their goal is to make a targeted service unavailable by overloading service provider resources with false requests. With resources depleted, the service provider is not able to serve legitimate users. Nowadays, DoS is a commonly-used attacking method which inflicts significant financial loss on its targets \cite{SKSS}. According to \cite{DM2004,MR2004} there are different types of DoS attacks. At the application level, attack detection is usually done by pattern recognition in the content of received packets. When a malicious pattern is detected, DoS prevention is achieved by blacklisting the IP address of the sender. To bypass this protection and to increase the efficiency of such attacks, attackers usually use distributed attacks (DDoS) by sending malicious packets from different source IP addresses, computers, networks or even continents. At present, detection of application-based attacks is very inefficient as a large number of packets has to be deeply inspected to recognize an attack pattern. We are tackling this problem at a much lower, network (or in some cases transport) layer, where deep packet analysis is not required. 

The motivation for this paper is twofold. On one hand, our goal is to propose an efficient method that combines the advantages of both feature based and volume based detection methods. On the other hand, we wanted to avoid the shortcomings of available datasets (see Section \ref{sec:related}), and that is why we decided to make a controlled DDoS experiment.

DDoS attacks are very simple to implement, which is the main reason why there is a wide spectrum of attacks in this category. Some of the most common DDoS attacks are ICMP flood \cite{LRST2000}, SYN flood  \cite{SKKSSZ1997}, DNS amplification attacks and the earlier Smurf Attack \cite{DD2015} and Fraggle Attack \cite{DD2015}. Most of these attacks aim to deplete either network or server resources.

In the following sections, we will describe the proposed detection method and the results of its evaluation. Section \ref{sec:related} describes the background of DDoS attacks and the related work in the area of DDoS attack detection. Section \ref{sec:method} describes the proposed detection method. Section \ref{sec:test} describes the testing scenario in which a real academic network is used. Section \ref{sec:Discussion} compares the results for different detection methods in the case of high-rate and low-rate DDoS attacks with best-case optimization for each method.

\section{Related Work}
\label{sec:related}

Mathematical modeling of a DDoS attack that would result in a practical, usable model (used for the provision of resources, etc.), is still an open issue, as DDoS attacks are changing. There are several approaches and we will mention only some of them. In \cite{WLLF2015}, Wang et al. model the system under SYN flood DDoS attack as a two-dimensional queuing model with N servers, two arrival processes and two service times with different distributions. Both the arrival of regular request packets and the arrival of attack packets are modeled as Poisson processes, but with different arrival rates $ \lambda _1$ and $ \lambda _2$. At most, N half-open connections are allowed at any one moment. A half-open connection for a regular request packet is held for a random time which is exponentially distributed. The two arrival processes are independent of each other and of the holding times for half-open connections. Based on these assumptions, DDoS is modeled as a two-dimensional embedded Markov chain. The authors give some security metrics for DDoS, such as a connection loss probability and buffer occupancy percentage of half-open connections for regular traffic.

Boteanu and Fernandez model SYN flood DDoS as an M/G/c/c birth-death Markov chain in \cite{BF2013}. 

There are several possible approaches to the testing of DoS detectors. The first, which is often used, is based on simulation. However, realistic simulation of network traffic is still an open issue which makes this method insufficient for reliable estimation of the detector performance. The second and best approach is to use a labeled real network dataset that contains both baseline and attack traffic. To the best of our knowledge, such publicly-available labeled datasets which include DDoS attack traffic and baseline traffic are very scarce. For example, there is a dataset that contains only attack traffic (CAIDA from 2007 \cite{CAIDA2007}) and there are datasets that contain only baseline traffic such as those collected by the MAWI working group \cite{WIDE}. The third approach is to use a public dataset that contains baseline traffic and to inject attack traffic synthetically. There are at least two issues with this approach. The first is the realistic generation of DoS traffic. The second is that network traffic is a coherent whole with interdependent segments. In real operation, when a server becomes overwhelmed with attack traffic, the dynamics of baseline packets are also changed. When the attack traffic is synthetically injected, this dependency is not present. 

Most of the authors in related works have tested their detectors using either simulated environments or publicly available traces like DARPA 1999 \cite{DARPA1999}, MIT KDD 99 \cite{KDD1999}, etc., most of which are obsolete today. None of those methods have been tested on real-life networks in real time. In some cases, normal traffic is sampled, and then attack packets are modeled and injected.
We have opted for a different approach, i.e., the fourth approach, which is to make a controlled DDoS experiment in a real network. Real-time access to packets in an academic network allows us to construct a detection method that is tested, calibrated and used in a real network. 

This paper is the continuation of work in \cite{BOP2014}. In \cite{BOP2014}, the authors have presented a comparison of two methods in the detection of SYN Flood attacks. One method is based on the number of SYN packets. It is presented by Siris and Papagalou in \cite{SP2004}. The other method is based on the entropy of packet addresses and on the CUSUM method. The conclusion in \cite{BOP2014} is that the entropy-based method achieves a performance close to that of the custom-tailored method, while on the other hand the entropy-based method is general and can detect a wide range of attacks and anomalies. Custom-tailored methods are often unusable for any type of attacks other than the one they have been designed for. In this paper, our aim is to test the findings from \cite{BOP2014}, but with real network traces.
 
CUSUM is actively used in both academic research and everyday practice – in many fields. Also, Ref \citep{SP2004} is widely cited. These two facts led us to the decision to use the method in \citep{SP2004} for comparison. 
 
DoS detectors are usually divided into volume-based and feature-based detectors. The experiments that we have performed resulted in the development of a hybrid method that is based both on packet entropy and packet rate, and that is a combination of volume-based and feature-based methods.

In the last decade, cloud systems entered everyday use and that fact highlights the importance of the research of DDoS attacks in the context of cloud systems, see \citep{GMDMM2017,GMDMR2017,GMDM2016}.

Our detector uses Shannon entropy, but other entropy formulas have also been used \cite{BOP2015,ZGMR2007,BJS2015}.

\section{The Proposed Detection Method}
\label{sec:method}
To be able to detect different types of DDoS attack we must focus on finding the common attributes of these attacks. One of these attributes is the diversity of source IP addresses that the attack packets come from. But relying on changes in diversity only can be misleading, as there are cases of legitimate use that result in such change. Typically, that is the case of short communications with multiple connections. Examples are a web page with a large number of external sources, and mail server distribution to a huge number of addresses, (e.g., a mailing list).

For that reason, our method uses the combination of diversity and packet rate: Exponential moving average (EMA) is applied to both values, using two EMA indicators – one uses short period (FastEMA), and the other one long period (SlowEMA) (Parameters values described in Table \ref{table:4emaparams}). The reasoning behind such an approach is explained in this section.

In our approach, the entropy of packet diversity H(z) is calculated using Shannon entropy equation (\ref{m:1}), where $z_{i}$ represents the number of packets exchanged between a communication pair (source IP address – destination IP address, or src-dst IP) and p($z_{i}$) represents the probability of occurrence calculated by dividing the total number of packets by $z_{i}$. 

\begin{equation}
\label{m:1}
H(z)=- \sum_{i=1}^n  z_{i} log(p( z_{i} )) 
\end{equation}
The entropy value is then normalized with the maximum entropy (\ref{m:2}) where N is the number of distinct communication pairs (src-dst IP). 
\begin{equation}
\label{m:2}
H(N)=log(N)/log(2)
\end{equation}

With the onset of a DDoS attack, the diversity, and subsequently the entropy will change. Whether it increases or decreases depends on the size of the network and the number of attackers.  

\begin{figure}[h]
\begin{center}
\includegraphics[width=\textwidth]{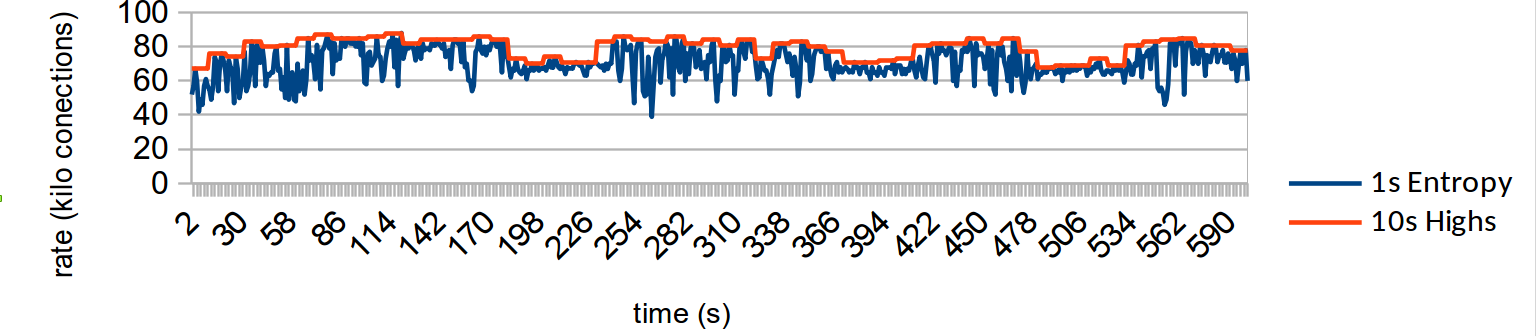} \caption{Signal with 10 second windows using the highest values}\label{fig:1stsignal}
\end{center}
\end{figure}

Real-life traffic is anything but regular with respect to packet diversity. Different services require different traffic rates, which creates irregular patterns of data flows. That is the reason why entropy values can vary greatly each second. Such variations prevent us from setting a threshold for DDoS attack detection. Entropy data must be filtered in such a way that only indicative values are used for decision making. Our method proposes filtering by grouping the entropy data in larger intervals of 10s, for example, and taking the highest entropy value as the representative value for each group (Fig. \ref{fig:1stsignal}). Such filtering is added to decrease packet variation sensitivity and only use the largest entropy and subset representative.
\begin{equation}
\label{m:3}
F(z)=max( z_{i},i=1..10)
\end{equation}

To avoid false alarms, the detection is further slowed down by using an exponential moving average algorithm (\ref{m:4}). Further information is given in \cite{K2011}.
\begin{equation}
\label{m:4}
\Delta  ^N_n =(1- \alpha )^{N-1}x_{n-N+1}+ \alpha  \sum_{s=0}^{N-2} (1- \alpha )^s x_{n-s}  	                                      
\end{equation}

where 
$  \alpha =\frac{2}{N+1}  $
\vspace{10pt}

EMA values are calculated on filtered entropy values to obtain the average entropy in a specific period. This technique successfully overcomes situations with short bursts of multiple connections but there are still problems with longer bursts that are not DDoS attacks. 

\begin{figure}[h]
\begin{center}
\includegraphics[width=\textwidth]{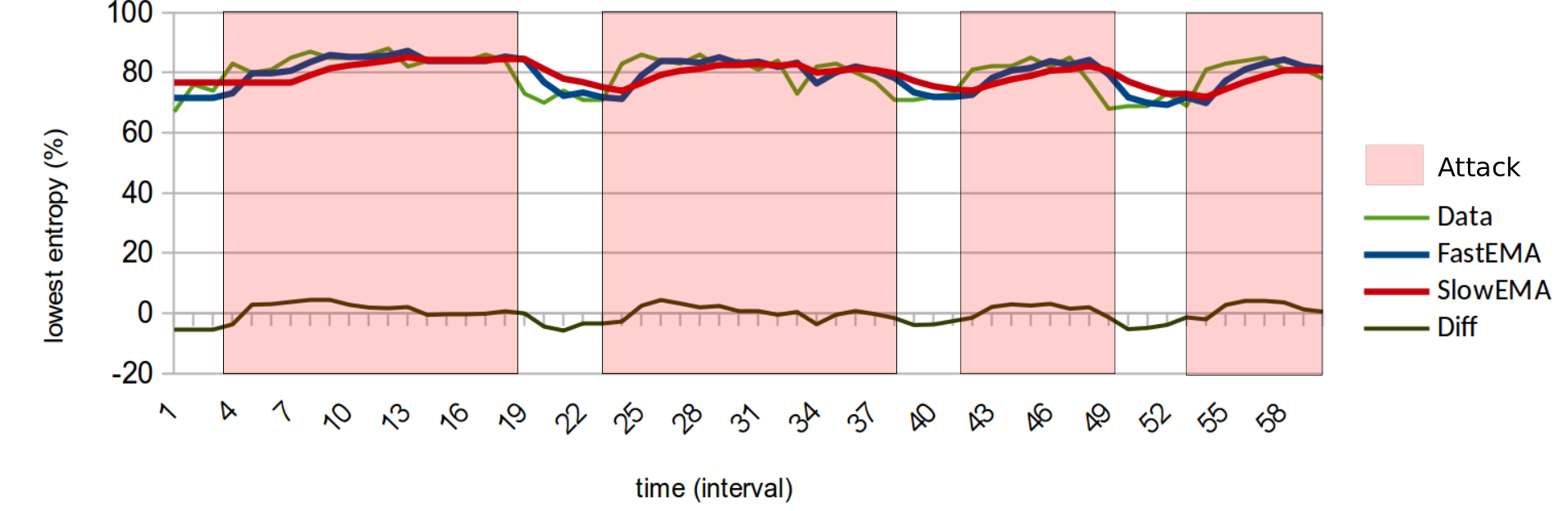} \caption{FastEMA, SlowEMA and Differential signal trends for entropy data}\label{fig:emadiff}
\end{center}
\end{figure}

To address the issue of bursts, and to set the decision-making procedure for automated DDoS detection, we propose the use of two EMA indicators for detection. One EMA indicator of filtered entropy values uses a short period, e.g., two samples. This is fast-changing EMA (FastEMA). The other EMA indicator uses a long period, e.g., six samples, which represents slow-changing EMA (SlowEMA). The two values show the fast and slow trends of the entropy changes respectively (Fig. \ref{fig:emadiff}). By calculating the difference between FastEMA and SlowEMA \ref{m:5} we can obtain a clear indication of the rising or dropping trend in filtered entropy values.

In our network testbed, a declining entropy trend can mean the beginning of a DDoS attack, as the packet diversity is increasing in set interval time (10s filtering * 2 samples = 20s for the FastEMA reaction). On the other hand, a rising trend can indicate the end of a DDoS attack.
\begin{equation}
\label{m:5}
diff(x)=FastEMA(x)-SlowEMA(x)
\end{equation}

For fine tuning of the decision-making algorithm, we have included thresholds for EMA differences. For example, the alarm is activated only if diff(x) is less than -1.2, and deactivated only if it is greater than 2. This optimization of the detector is possible in networks with a specific traffic profile.
The proposed procedure can be optimized for the detection of a large spectrum of DDoS attacks. But entropy detection is based only on the packet diversity. In certain cases, the utility of such a detector is limited. Consequently, we cannot base our method solely on one common attribute of DDoS attacks. In addition to packet diversity, we decided to use packet rate as another common attribute of attacks.

\begin{figure}
\begin{center}
\includegraphics[width=\textwidth]{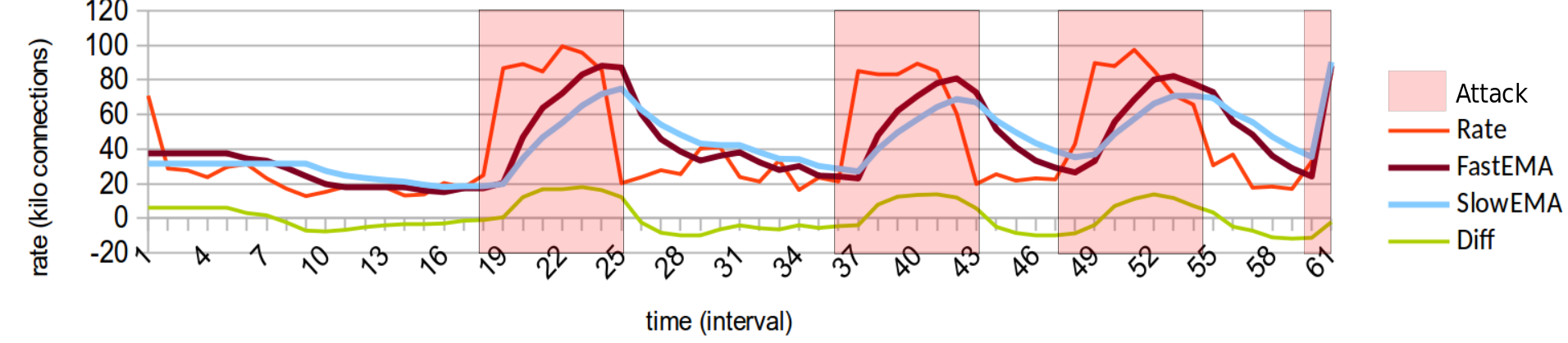} \caption{FastEMA, SlowEMA and Differential signal trends for packet rate data}\label{fig:emadiffpkt}
\end{center}
\end{figure}

Using exactly the same procedure as described for entropy values, we can calculate the EMA difference of packet rates. With a period of one second, and a sliding window of 10 seconds, the representative with the highest value is chosen. Based on these samples, the FastEMA is calculated using four samples that have packet rate values with higher variance than the entropy. The SlowEMA uses an interval of eight samples. The difference between FastEMA and SlowEMA is then calculated, and the threshold is applied to fine-tune the detection (Fig. \ref{fig:emadiffpkt}).

\begin{lstlisting}[ basicstyle=\ttfamily\scriptsize, language=Java, caption=4EMA pseudo-code,label={lst:code1}]
struct pair {
src: IP
dst: IP
counter: int
}

double currentEntropyData[], currentRateData[]
double filteredEntropyData[], filteredRateData[];

pair packets[]=currentSecondPackets();

double current_entropy=packetEntropy(packets);
currentEntropyData[]=current_entropy;
double current_rate=packetRate(packets);
currentRateData[]=current_rate;

double fent=max(currentEntropyData, interval);
fltEntropyData[]=fent;
double frate=max(currentRateData, interval);
fltRateData[]=frate;

double entFastEma=EMA(fltEntropyData, emaFastInt);
double entSlowEma=EMA(fltEntropyData, emaSlowInt);
double rtFastEma=EMA(fltRateData, emaPacketFastInt);
double rtSlowEma=EMA(fltRateData, emaPacketSlowInt);

double entDiff=entFastEma-entSlowEma;
double rtDiff=rtFastEma-rtSlowEma;

if (entDiff < trEntAlarm && rtDiff > trPktAlarm)
	alarm=1;
if (entDiff > trEntNoAlarm && rtDiff < trPktNoAlarm)
	alarm=0;

\end{lstlisting}

\begin{table*}[h!]
\centering
\normalsize
 \begin{tabular}{||c c c||} 
 \hline
 \textbf{Parameter} & \textbf{Typical value} & \textbf{Description} \\ [0.5ex] 
 \hline\hline
 emaFastInterval & 2 samples (20 seconds) & Fast EMA interval for entropy data\\ 
 \hline
 emaSlowInterval & 4-6 samples & Slow EMA interval for entropy data \\
 \hline
 trEntAlarm & -0.74 & Threshold for entropy EMA Diff signal  \\
 \hline
 trEntNoAlarm & 0.10 & Threshold for entropy EMA Diff signal that indicate end of attack\\
 \hline
 emaPacketFastInterval & 4 samples & Fast EMA interval for packet rate data\\ 
 \hline
 emaPacketSlowInterval & 8 samples & Slow EMA interval for packet rate data\\ 
 \hline
 trPktAlarm & 0.10 & Threshold for rate EMA Diff signal\\ 
 \hline
  trPktNoAlarm & -0.50 & Threshold for rate EMA Diff signal that indicates end of attack\\ 
 \hline
\end{tabular}
\caption{4EMA parameters}
\label{table:4emaparams}

\end{table*}

\begin{figure}
\begin{center}
\includegraphics[width=12cm]{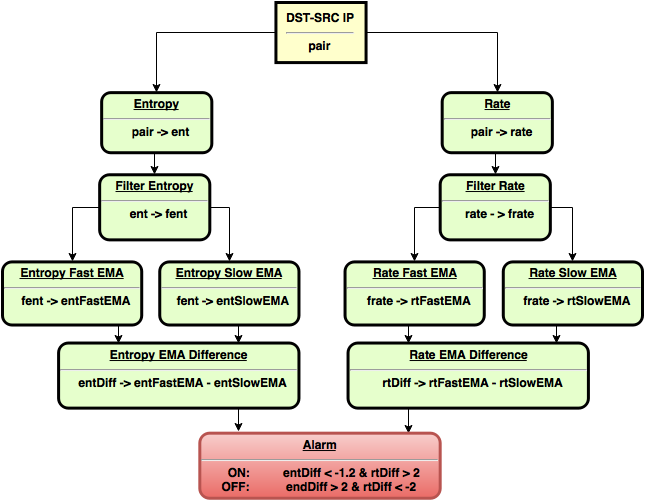} \caption{4EMA algorithm scheme}\label{fig:4emadiagram}
\end{center}
\end{figure}

Thus, thresholds are applied to both entropy EMA difference and rate EMA difference. Fig. \ref{fig:4emadiagram} and pseudo-code (Listing \ref{lst:code1}). presents the complete algorithm with the parameters described in Table  \ref{table:4emaparams}. The parameter values, in the Table \ref{table:4emaparams}., are generated by both automatic optimization process and manual tuning with the aim to get the best possible result in different tested attack samples.

For easier identification, the proposed method is henceforth referred to as 4EMA in the text.

\section{Test Scenario}
\label{sec:test}

\begin{figure*}[h!]
\begin{center}
\includegraphics[width=\textwidth]{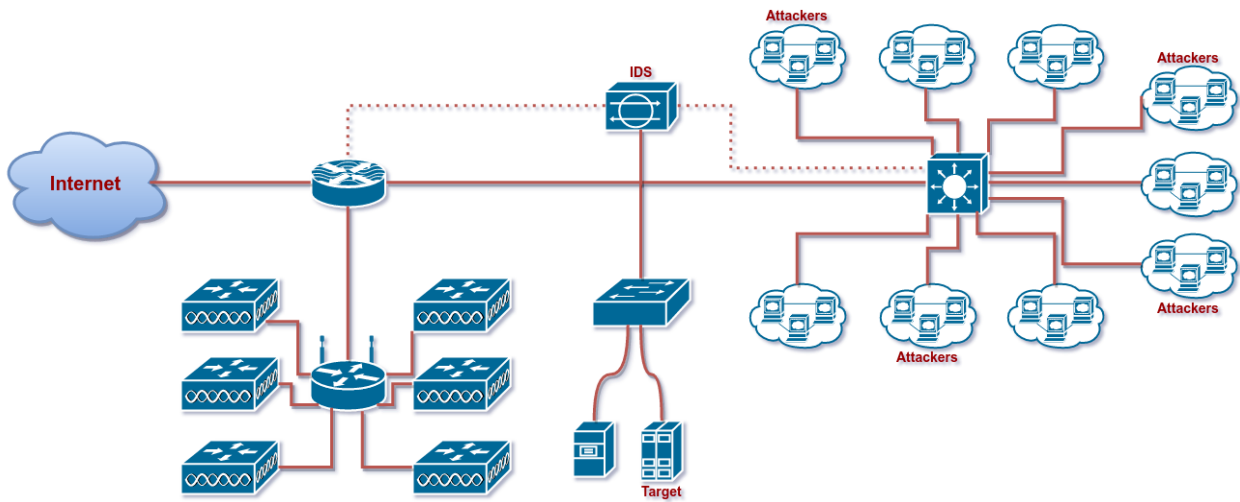} \caption{Network diagram for test scenario}\label{fig:testscenario}
\end{center}
\end{figure*}

Simulation of computer networks reproduces specific scenarios of network operation. Usually, these scenarios are based on a particular case and evaluation is thus limited to that specific case. The most important challenge for any method is to be tested in a real-life situation where traffic is more unpredictable.

Our method has been tested in a real-life computer network environment (Fig. \ref{fig:testscenario}). For testing, we used an academic computer network which included about 300 students’, professors’ and employees’ workstations and about 20 servers.
For the purpose of the test, we developed a prototype intrusion detection system (IDS) able to collect traffic coming from the network, to calculate variables specified by the proposed method, to make alarm decisions and to store variables in a data log file for further analysis. The prototype IDS is hosted by one of the servers. The central network router is configured with CALEA tools (Communications Assistance for Law Enforcement Act) to redirect a sample of every packet to the IDS.

The academic network uses a Mikrotik router, so data are encapsulated in TaZmen Sniffer Protocol (TZSP). The IDS listens to TZSP UDP port 37008, de-encapsulates forwarded data and extracts network and transport layer information for each packet. After extracting the IP header data for each communication pair (src-dst IP), the number of packets is counted until the sample interval has expired. Information about protocol and TCP flags is also stored.

At the end of each sample interval, collected data for the communication pair are stored in the data log file, the counter is reset and the proposed method of detection is applied.

As part of the testing scenario, we have created a baseline by collecting data from the network when an attack is not present. The analysis of that traffic confirms that the entropy values in the case of normal operation (with no active attack) are not stable, which eliminates the possibility of using simple threshold detectors. 
     
To test the performance of the proposed method we set up a botnet network inside the university network (Fig. \ref{fig:testscenario}). Using the Microsoft Windows Active Directory service, we distributed the attack script file to some of the computers in students’ classrooms. The intensity of the conducted attack was chosen carefully so as not to harm the normal operation of the university network.

A script file was developed to test the following attacks:
\begin{itemize}
\item ICMP flood attack with a high packet rate attack on the specified target. The packets are sent at maximum speed.
\item TCP SYN attack that sends a configured number of SYN packets per second to specified targets. This attack is realized by sending a large number of SYN packets to the attack target. The target server responds by sending SYN-ACK packets, but attackers do not complete the TCP three-way handshake, i.e., they do not send ACK packets. Thus, the resources allocated for connection establishment are not released until the timeout.
\end{itemize}

While the ICMP flood attack aims to deplete the victim’s network resource and is a representative of a network resource intensive attack, TCP SYN is representative of a moderate network intensity attack which targets other types of resources (such as CPU, RAM, etc.). In the case of the SYN flood, the attacker tries to exhaust the so-called backlog of half-open connections associated with a port number. The backlog is a system limit on the number of TCP Control Block (TCB) structures that can be resident at any time \cite{RFC4987}. Although we have not tested UDP flood attack, its effects are expected to be similar to the two we have tested. 

The script that is distributed to computers stays resident in the memory after a student logs in, and it monitors the specific target IP address. When the target IP address becomes reachable, (i.e., it responds to ICMP messages), all active computers in this botnet begin the attack on the target. If it is not reachable, the check is repeated after ten seconds.

Using this approach, a real-life botnet is simulated. Usually, the botnet master stores instructions at a fixed location and bots in the botnet periodically check for the existence of these instructions. When instructions are found, the attack begins. So, as in a real-life situation, the attack does not start with all attackers at the same time, but instead with attackers initiated after a random delay period (maximum 10s in our example).

By monitoring network conditions and by taking into account the schedule of network operations, we have chosen the moment to initiate the attack. At that moment, the IP address that the botnet script monitors is assigned to one of the dummies (honeypot) computers. Within the 10s interval, the botnet computers begin continuous attacks that last as long as that IP address is present in the network. Not all computers are part of the botnet. The Active Directory is used to distribute the script to only some of the computers in the students’ classrooms, and only computers with logged-in students execute the script. This creates a very variable environment when, during testing, computers are powered on or off and students are logged in and out, and this results in an attack with inconsistent intensity as in a real-life DDoS attack.

The characteristics of one of the ICMP flood attack tests are given in Table \ref{table:icmpflood}.

\begin{table}[h!]
\centering
\normalsize
\caption{ICMP flood attack statistics}
\label{table:icmpflood}
 \begin{tabular}{||c c||} 
 \hline
 \textbf{ICMP Packets per second per attacker} & 1000 \\ 
 \hline
 \textbf{Number of attackers during attack} & 33-42 \\ 
 \hline
 \textbf{ICMP packet size} & 70 bytes \\ 
 \hline
 \textbf{Attack duration} & 1800 seconds \\ 
 \hline
 \textbf{Per attacker attack rate} & ~70 kbps  \\ 
 \hline
 \textbf{Cumulative traffic rate during attack} & 47-390 kpps \\ 
 \hline
\end{tabular}
\end{table}

When testing the detection of a TCP SYN attack we can control the number of SYN packets sent by the attacker which allows us to test our method in different intensity cases. To avoid the results overlapping with the ICMP flood attack we skipped the test with the maximum number of TCP SYN packets that the attacker can generate, and concentrated on lower-rate attacks, as they are more difficult to detect. The first round of TCP SYN tests is initiated with an attack rate of 100 packets per second (per attacker).

\section{Discussion of Results}
\label{sec:Discussion}
Most high-rate DDoS attacks, e.g., ICMP or UDP flood attacks aim to deplete the server’s network resources by sending a large number of packets. These types of attacks can be detected even with a packet rate indicator as a detection mechanism. On the other hand, regular traffic can have the same fingerprint, e.g., a server with Internet radio content that generates a large number of small UDP packets, or a server that provides a huge number of images for an image gallery. Combining entropy change detection with packet rate change increases the reliability of DDoS detection.

\begin{figure}[h!]
\begin{center}
\includegraphics[width=\textwidth]{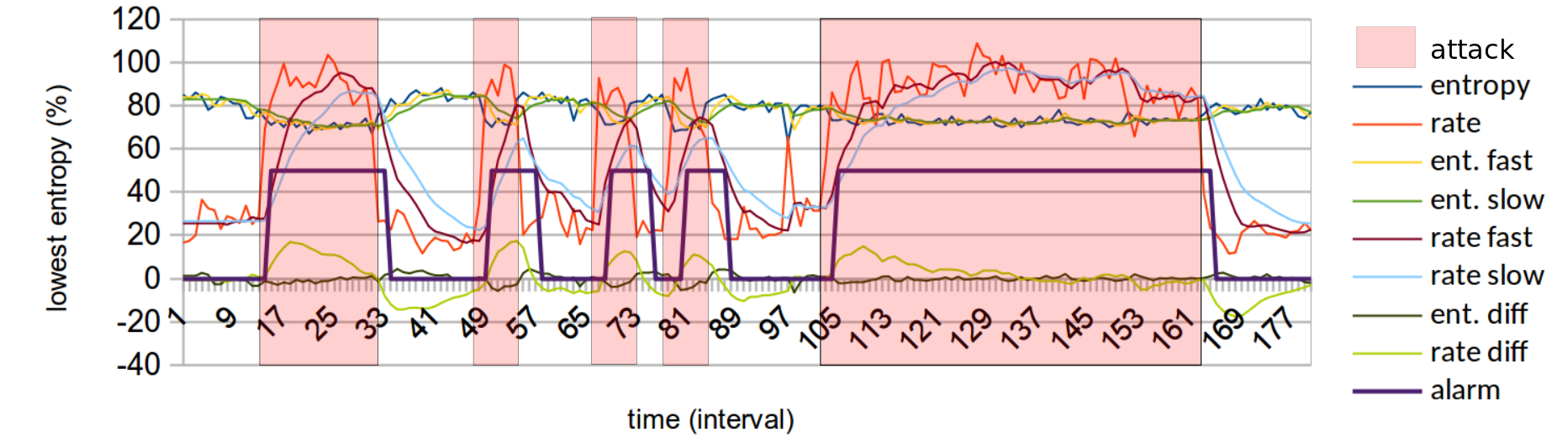} \caption{4EMA high-rate attack detection}\label{fig:highrateattack}
\end{center}
\end{figure}

Fig. \ref{fig:highrateattack} shows the performance of the proposed detection method with high-rate attacks (ICMP flood attack, Table 1). In addition to the indication in the form of an increased packet rate, the figure shows increased packet diversity (a decreased entropy factor). By considering the difference between slow and fast trends of both the rate and the entropy we can set up thresholds for the IDS alarm. 

In the 99th period of the figure, a packet burst has occurred that is part of the regular traffic. By using slow EMA trends for both indicators, we can ignore short bursts with similar fingerprints to attacks. The use of slow trends also slows down the detection of attacks (increases the detection delay). Therefore, the minimum detection delay is 60s as this is the interval for the slowest EMA. Decreasing this delay by decreasing the interval for slow trends can result in an increase in false positives for regular traffic. 

The proposed method detected all the tested high-rate attacks. With the optimization of parameters, the proposed method resulted in no false positives. The optimized parameters were then tested with real-time traffic without introducing the attacks. The purpose of this test was to examine the occurrence of false positives. During the 12 hours of the test, we obtained only a few false positives in the two following cases:

\begin{itemize}
\item if traffic bursts take longer than the slowest trend (60s),
\item with the use of BitTorrent or similar peer-to-peer (P2P) protocols.
\end{itemize}

The next step was to test the detection of low-rate attacks. These attacks present a true challenge for detection at the network layer. Attacks of 100 or fewer packets per second are considered low-rate attacks as they usually consume a very small amount of bandwidth. Our test setup has enabled us to examine the performance of algorithms in the case of low-rate attacks also.

\begin{figure}[h!]
\begin{center}
\includegraphics[width=\textwidth]{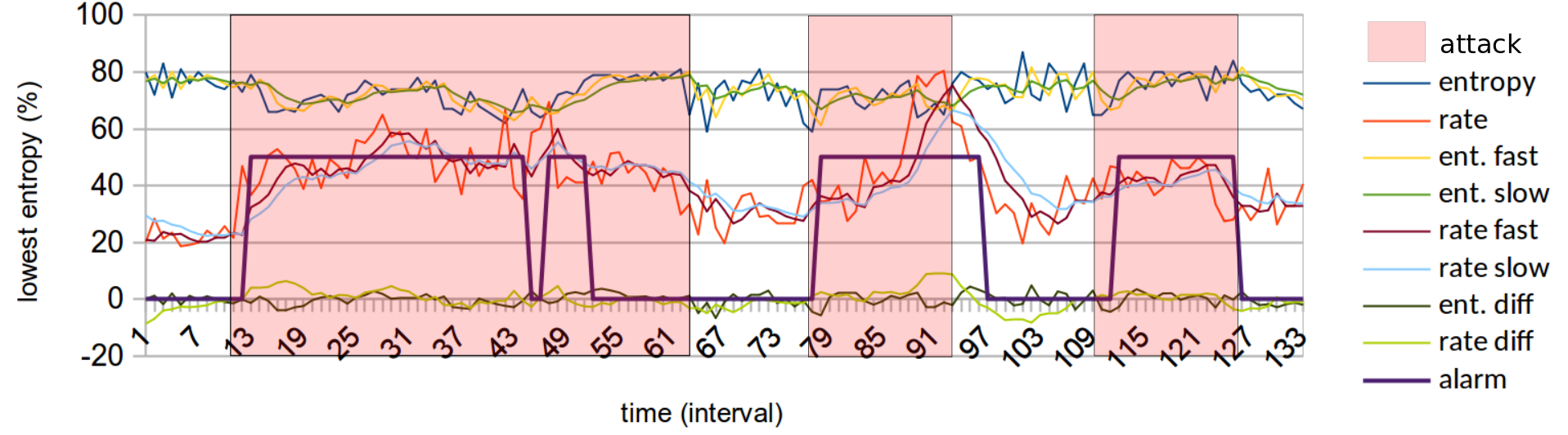} \caption{4EMA low-rate attack detection }\label{fig:lowrateattack}
\end{center}
\end{figure}

Fig. \ref{fig:lowrateattack} shows the results in the case of a 100 pps TCP SYN flood attack. Results show that the proposed detection method can be used even with low-rate attacks, with a high level of reliability. Without further optimization, this method detected all attacks. However, it reported more alerts than actual attacks. In the case of one attack, it alerted twice, incorrectly assuming that the attack had stopped and another attack had started.  On the other hand, the detector resulted in no false positives in the tested data range.

\begin{figure}[h]
\begin{center}
\includegraphics[width=\textwidth]{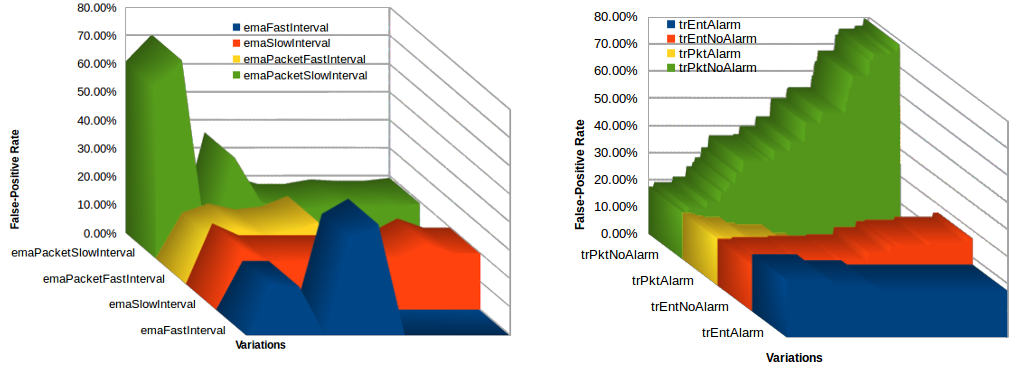} \caption{4EMA False-Positive rate parameters variation
 }\label{fig:fpr}
\end{center}
\end{figure}

\begin{figure}[h]
\begin{center}
\includegraphics[width=\textwidth]{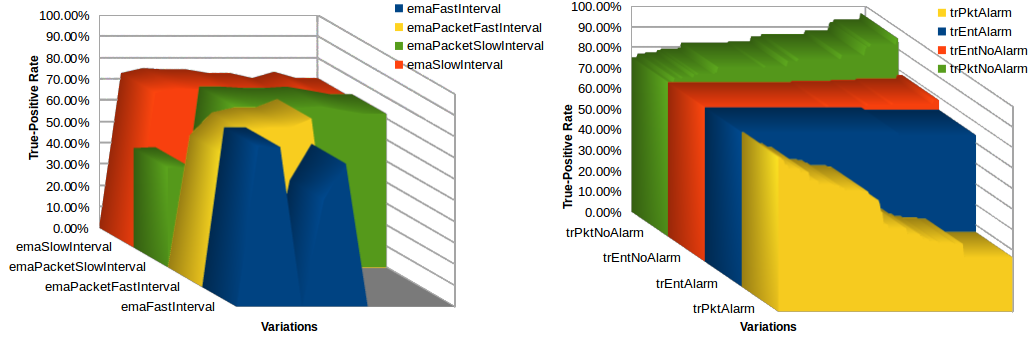} \caption{4EMA True-Positive rate parameters variation
 }\label{fig:tpr}
\end{center}
\end{figure}

Depending on optimization, 4EMA algorithm can provide a flexible way of detecting both high-rate and low-rate attacks. Figure \ref{fig:fpr} represents diagram of False-Positive detection rate with variation of 4EMA parameters described in Table \ref{table:4emaparams}.
 FastEMA intervals (emaFastInterval, emaPacketFastInterval), for both Entropy and Packet indication, greatly influence False-Positive rate. Thresholds parameters are only there to fine-tune Alarm and No-Alarm states. Figure \ref{fig:tpr}, displays diagram for parameters optimization of True-Positive rate. Even if a variation of SlowEMA parameters does not make a huge difference in detection, careful optimization between Fast and Slow EMA is required for the best detection performance, before it is fine-tuned with Thresholds. For detection rates, data were sampled using attack described in Table \ref{table:synflood}.

We have compared the results of our 4EMA algorithm with the CUSUM-SYN \cite{SP2004} and CUSUM-Entropy \cite{BOP2015} detection methods. 

The test scenario includes multiple attacks and the response is measured for each detection method with dataset \cite{BBSM}. We are using the same input dataset for all detection methods, so we can compare their results. High-intensity ICMP flood attacks cannot be detected by the CUSUM-SYN method, so we cannot compare the two methods with respect to this type of attack. The scenario includes an SYN Flood attack as described in test scenario section with attack parameters given in Table \ref{table:synflood}.

\begin{table}[h!]
\centering
\normalsize
\caption{SYN Flood attack statistics}
\label{table:synflood}
 \begin{tabular}{||c c||} 
 \hline
 \textbf{TCP SYN packets per second per attacker} & 20 \\ 
 \hline
 \textbf{Number of attackers during attack} & 55-60 \\ 
 \hline
 \textbf{TCP SYN Flood attack packet size} & 60 bytes \\ 
 \hline
 \textbf{Attacks testing duration} & 60 minutes \\ 
 \hline
 \textbf{Attack rate per attacker} & ~1.2 kbps \\ 
 \hline
 \textbf{Cumulative traffic rate during attack} & 1100-1200pps\\ 
 \hline
 \textbf{Number of attacks} & 21 \\ 
 \hline
 \textbf{Individual attacks duration} & 60-180s \\ 
 \hline

\end{tabular}
\end{table}

\begin{figure}[h]
\begin{center}
\includegraphics[width=\textwidth]{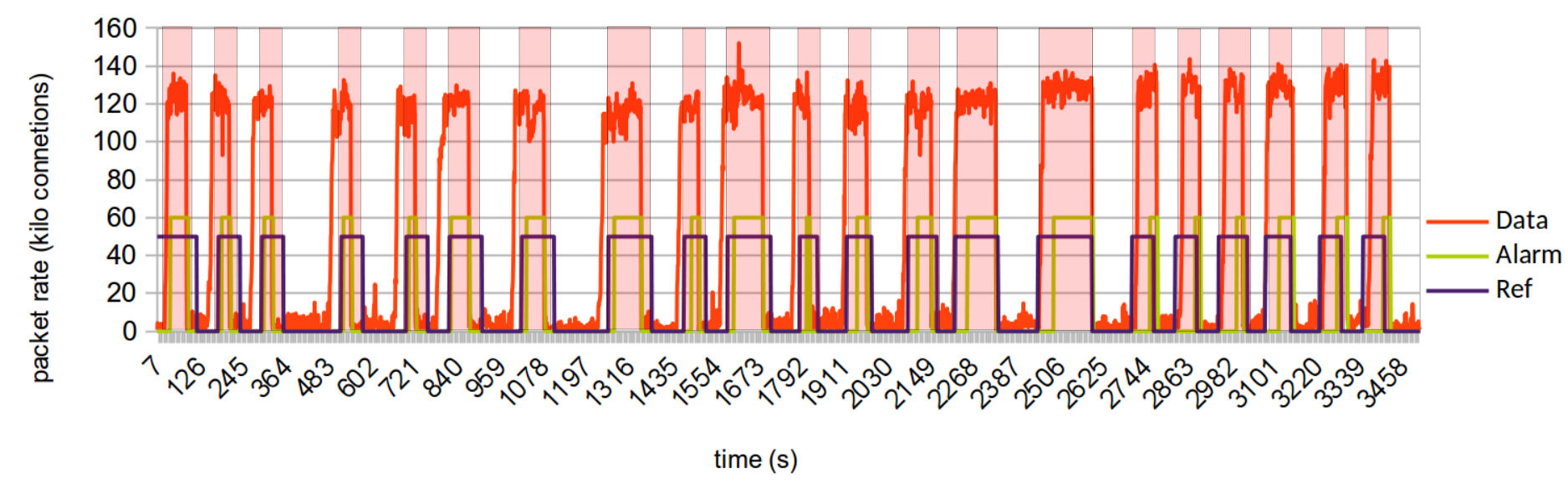} \caption{CUSUM-SYN detection}\label{fig:cusumsyn}
\end{center}
\end{figure}

Fig. \ref{fig:cusumsyn} presents the graph of CUSUM-SYN detection with optimized parameters (β1=0.148, β2=3, k=18, h=6.8, K=0.01). The Data signal represents the total number of filtered SYN packets which is the input for CUSUM-SYN detection. The Alarm signal represents the result of the CUSUM-SYN detection method and the Ref signal is the reference for the attacks. With filtration of TCP SYN packets, detection of attacks is achieved relatively easily, as there is a huge difference between the signals observed when the attack is present and when it is not. From the results, we can confirm that CUSUM-SYN detected all the TCP SYN attacks without any false negatives.

\begin{figure}[h]
\begin{center}
\includegraphics[width=\textwidth]{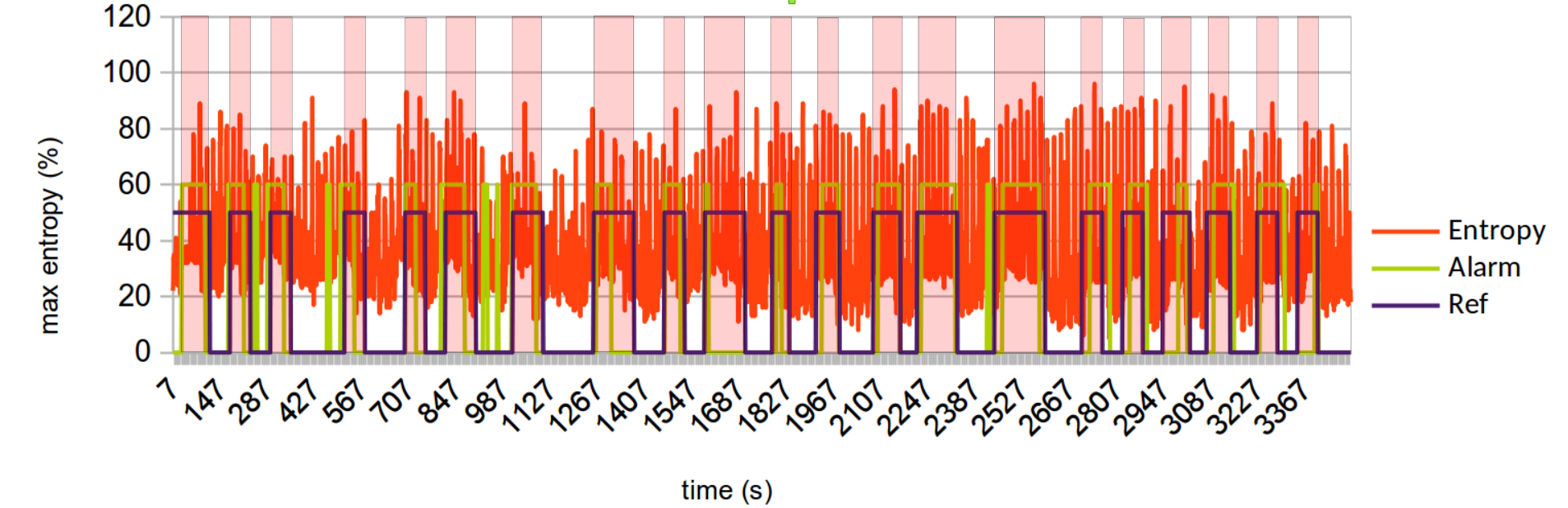} \caption{CUSUM entropy detection}\label{fig:cusumentropy}
\end{center}
\end{figure}

Fig. \ref{fig:cusumentropy} presents the graph of CUSUM entropy detection which uses the entropy of all packets to detect an attack. The results are provided for optimized parameters (β1=0.139, β2=0.0001, k=28, h=3.1, K=0.51). The Alarm signal represents the result of this method of detection. Using Ref as a reference signal, we can conclude that this method detected all attacks in this scenario but that there are also some false positives (about 14\% as shown in Table \ref{table:detectionmethods}). Despite a slightly lower reliability, CUSUM entropy detection has one important advantage compared to CUSUM-SYN, and that is the ability to detect other types of DDoS attacks, apart from TCP SYN attack. 

\begin{figure}[h]
\begin{center}
\includegraphics[width=\textwidth]{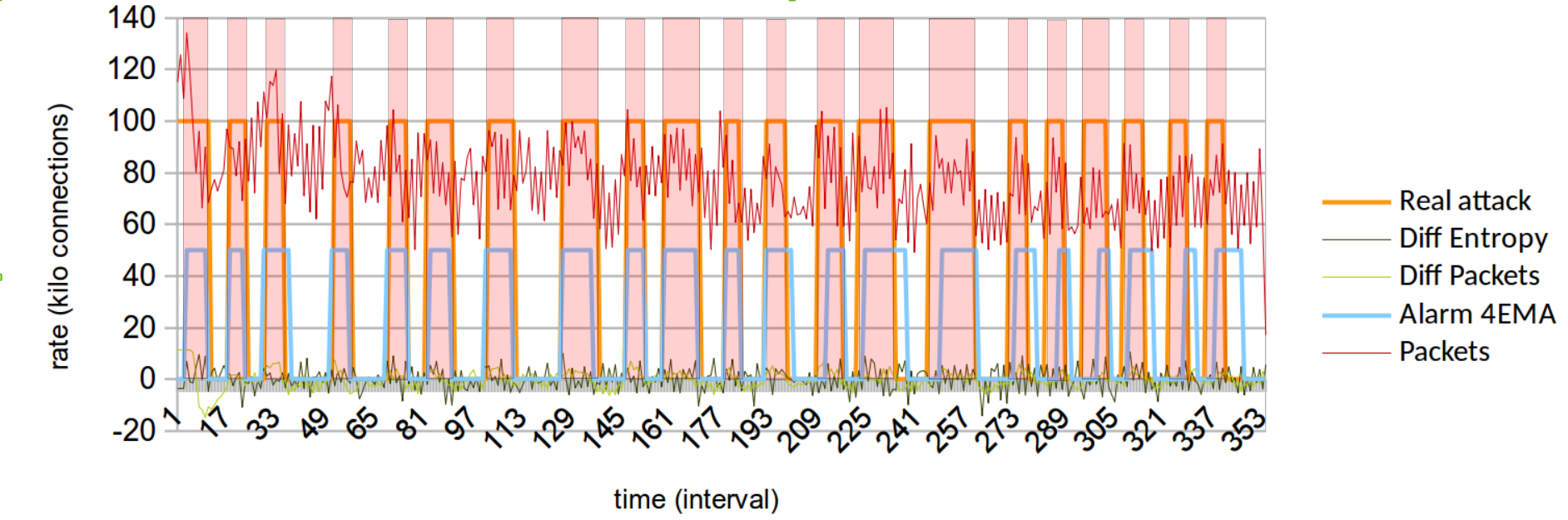} \caption{4EMA entropy detection}\label{fig:4emaentropy}
\end{center}
\end{figure}

Fig. \ref{fig:4emaentropy} shows the performance of the proposed detection method. The Real attack signal is our reference signal, and Diff entropy and Diff packets are the differential components of 4EMA detection. The alarm is the result of 4EMA detection. Packets indicator is the signal that represents the total number of all packets divided by 10. So, the total packet rate in this test case is between 500 and 1300 packets per second (the attack intensity is still only 120 pps). 

The proposed method is optimized for this test case (emaSlowInterval=4.0, emaFastInterval=2.0, trPktAlarm=0.01, trPktNoAlarm=0.5, emaPacketFastInterval=4.0, emaPacketSlowInterval=8.0, trEntAlarm=0.74, trEntNoAlarm=0.1). The results of the test case show reliable detection of all attacks without false positives. By using more than just the entropy of all packets, this detection method is more reliable than CUSUM Entropy and in addition, it is designed to detect not just TCP-based attacks (contrary to CUSUM-SYN), but also attacks with similar characteristics that are common to any DDoS attack.

\begin{figure}[h]
\begin{center}
\includegraphics[width=12cm]{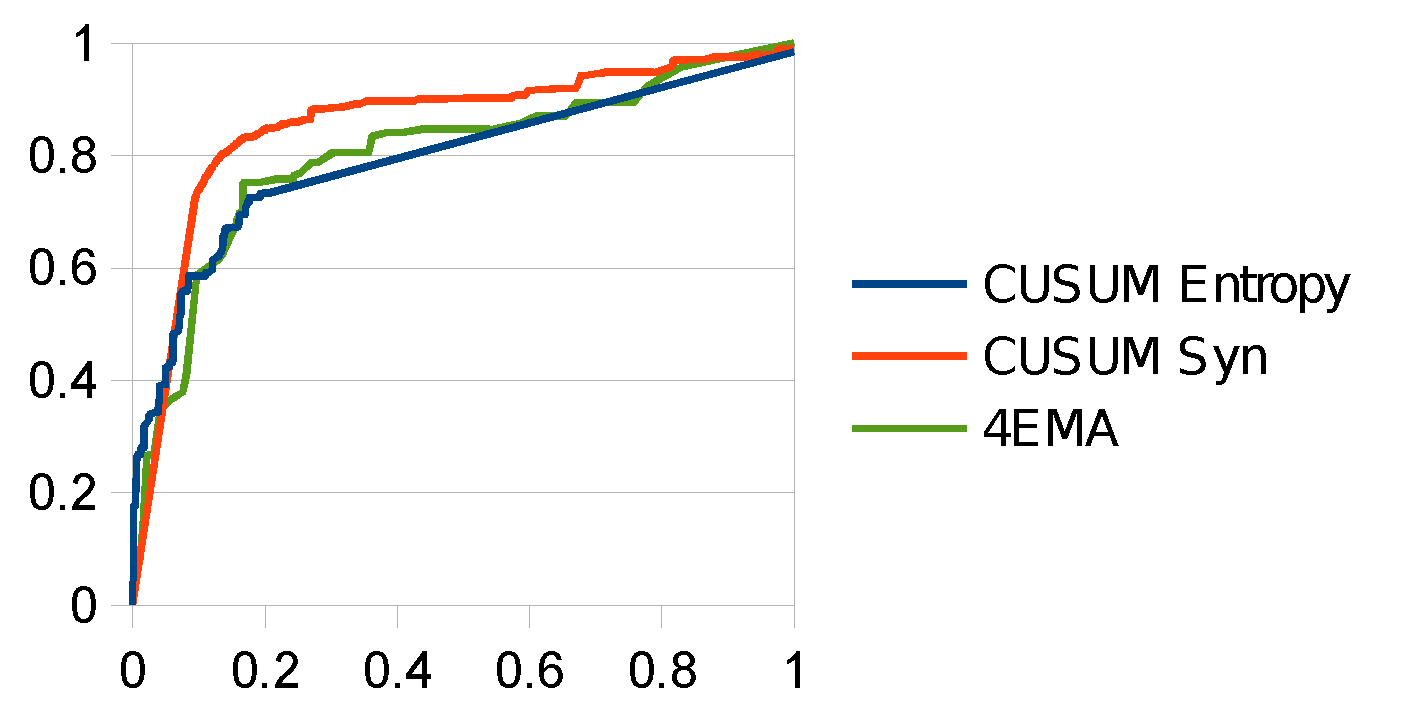} \caption{RoC curve for detection methods parameters
 }\label{fig:roc}
\end{center}
\end{figure}

The results are presented for the case with optimized parameters. The variation of detection results can also be displayed with a receiver operating characteristic (RoC) curve for each detection method (Fig. \ref{fig:roc}).

From the figure above, we can see that CUSUM-SYN is more reliable and less optimization sensitive than the other two detection methods. This is simply because it uses only filtered TCP SYN packets so there is less room for error. The other two detection methods have similar parameter sensitivity as they both use non-filtered data for attack detection.

With respect to the complexity of the three algorithms, they all scale as O(N), where N is the number of packets in the observed interval. 

We have calculated the F1 score for tested methods. The best score for each method is presented in Table \ref{table:f1score}.

\begin{table}[h!]
\centering
\normalsize
\caption{F1 score for tested methods}
\label{table:f1score}
 \begin{tabular}{||c c c c||} 
 \hline
 \textbf{Detection method} & \textbf{Recall} & \textbf{Precision}& \textbf{F1 score} \\ 
 \hline
 CUSUM-SYN & 0.76 & 0.86 & 0.81 \\ 
 \hline
 CUSUM Entropy &0.59 & 0.86 & 0.70 \\ 
 \hline
 4EMA Entropy & 0.76 & 0.90 & 0.82 \\ 
 \hline
\end{tabular}
\end{table}

\begin{figure}[H]
\begin{center}
\includegraphics[width=12cm]{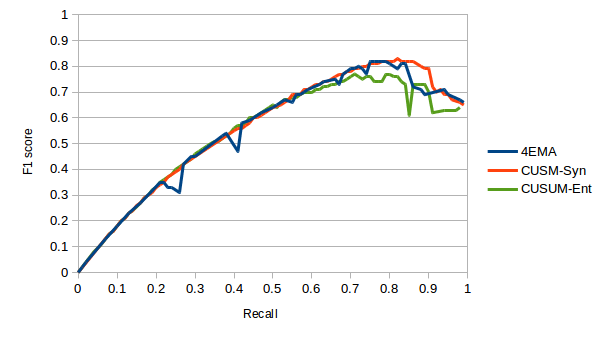} \caption{F1 score in correlation with recall for every parameter variation
 }\label{fig:F1score}
\end{center}
\end{figure}

To provide better awareness of dependency between parameters variation and F1 score we are showing, in  Fig. \ref{fig:F1score}, a graph of the F1 score in a correlation of calculated recall.

\section{Conclusion}
\label{sec:conclusion}
The contribution of this paper is twofold. Firstly, a hybrid approach for detection of distributed denial-of-service attacks is proposed. Secondly, the performance of the proposed approach has been evaluated and compared to two existing methods. Therefore, a controlled DoS experiment in a real network has been realized to generate the data sets used for evaluation. The experiment included two types of attacks: an ICMP flood attack, representing a high-rate attack and a TCP SYN flood attack, representing a low-rate attack. 

The proposed approach is based on application of the exponential moving average to two time series, one containing entropy values and the other containing numbers of packets. Thus, the method incorporates both volume-based and feature-based detection. We use two EMA indicators for both time series – one with a short period and the other with a long period. To evaluate the performance, several indicators have been used: detection rate, recall, precision and F1 score. The receiver operating curve has also been provided.

One of the methods used for comparison is a custom-tailored method for detection of SYN flood attacks and the other is a general entropy-based anomaly detection method.

\begin{table*}[h]
\centering
\normalsize
\caption{Comparison of detection methods}
\label{table:detectionmethods}
 \begin{tabular}{||c c c c||} 
 \hline
 \textbf{Detection method} & \textbf{\makecell{Detection\\rate}} & \textbf{\makecell{False\\Positives}} & \textbf{  \makecell{Able to detect \\UDP/ICMP\\attacks}} \\ [0.5ex] 
 \hline\hline
 CUSUM-SYN \cite{SP2004} & 100\% & 0\% & No\\ 
 \hline
 CUSUM Entropy \cite{BOP2015} & 100\% & ~14\% & Yes \\
 \hline
 4EMA Entropy [Section \ref{sec:method}] & 100\% & 0\% & Yes  \\
 \hline
\end{tabular}

* All parameters of the method are optimized for their best performance on the tested scenario

\end{table*}

The results of the comparison of the three detection methods are given in Table \ref{table:detectionmethods}. We see that the proposed method achieves the detection rate of the custom-tailored detection method, but on the other hand, it has generality with regard to detection of multiple types of attacks [Table \ref{table:summary}].

\begin{table*}[h]
\centering
\normalsize
\caption{Summary of 4EMA tests with different attacking intensities}
\label{table:summary}
 \begin{tabular}{||c c c||} 
 \hline
 \textbf{} & \textbf{\makecell{High-rate\\ attack}} & \textbf{\makecell{Low-rate\\ attack}} \\ 
 \hline
 \textbf{Attack rate per attacker} & 1000 pps & 20 pps \\ 
 \hline
 \textbf{Cumulative attack packet rate} & 47-390 kpps & 1000-1200 pps \\ 
 \hline
 \textbf{Cumulative traffic rate} & 3.6-29mbps & 6-70kbps \\ 
 \hline
 \textbf{Number of attackers} & 33-42 & 55-60 \\ 
 \hline
 \textbf{Attacking methods} & \makecell{ICMP,\\UDP Flood} & \makecell{TCP Syn\\ICMP\\UDP Flood} \\ 
 \hline
 \textbf{Detection rate} & 100\% & 100\% \\ 
 \hline
 \textbf{False positives} & 0\% & 0\% \\ 
 \hline

\end{tabular}

\end{table*}

One shortcoming of the method is its inability to distinguish denial-of-service traffic from peer-to-peer traffic, (e.g., BitTorrent). That is the common drawback of a majority of existing methods, and it represents one of the directions of our future work. The other is the extension of the proposed algorithm with a preprocessing module that would enable its application in congested links connecting large networks.

\section*{Acknowledgements}
This research was financially supported by the Ministry of Education, Science and
Technological Development of the Republic of Serbia through Projects No. III 45003 and III 44009-2.

\section*{References}

\bibliography{reference}

\section*{Biographies}
\noindent {\bf Petar D. Bojovic} has graduated with Master degree on Faculty of Computer science on Union University Belgrade in 2008. In June 2008, he joins Faculty of Computer science as Lecturer in the department of computer networks.  Presently he works as Associate Professor at Faculty of Computer science on teachings and research. His interest includes Computer networks and Security of computer networks.

\noindent {\bf Ilija Basicevic} received Dipl. Eng. M.Sc, and Ph.D. degrees from the University of Novi Sad. Currently, he is the associate professor at the University of Novi Sad, teaching courses on computer networks. His research interests are in the areas of Internet protocols and network security. He has authored or co-authored more than 45 scientific papers and one textbook.

\noindent {\bf Stanislav Ocovaj} received his B.Sc. degree in electrical and computer engineering in 2004, and his M.Sc. degree in 2010 from the Faculty of Technical Sciences at the University of Novi Sad, Serbia.

\noindent {\bf Miroslav  V. Popovic} received his Dipl. Eng., M.Sc., and Ph.D. degrees from the University of Novi Sad, Serbia. Currently, he is the full professor at the University of Novi Sad. His research interests are system programming, distributed systems, and security. He has published about 20 journal papers, more than 120 conference papers and the book Communication protocol engineering (CRC Press).

\end{document}